\documentclass[letterpaper,twocolumn,10pt]{article}
\usepackage{usenix2021}
\usepackage{amsmath}
\usepackage{booktabs}
\usepackage{balance}
\usepackage{graphicx}
\usepackage{xcolor}
\usepackage{url}
\usepackage{hyperref}
\usepackage{listings}
\usepackage{tikz}

\definecolor{mGreen}{rgb}{0,0.6,0}
\definecolor{mGray}{rgb}{0.5,0.5,0.5}
\definecolor{mPurple}{rgb}{0.58,0,0.82}
\definecolor{backgroundColour}{rgb}{0.95,0.95,0.92}

\lstdefinestyle{CStyle}{
    commentstyle=\color{mGreen},
    keywordstyle=\color{magenta},
    numberstyle=\tiny\color{mGray},
    stringstyle=\color{mPurple},
    basicstyle=\footnotesize,
    breakatwhitespace=false,         
    breaklines=true,                 
    captionpos=b,                    
    keepspaces=true,                 
    numbers=left,                    
    numbersep=5pt,                  
    showspaces=false,                
    showstringspaces=false,
    showtabs=false,                  
    tabsize=2,
    language=C
}

\usetikzlibrary{tikzmark}
\tikzset{mycircled/.style={circle,draw,inner sep=0.1em,line width=0.04em}}

\def\ocirc#1{\tikzmarknode[mycircled,draw=orange,fill=orange]{t1}{#1}}

\def\BEGINITEMIZE{\begin{itemize}}
\def\BEGINENUMERATE{\begin{enumerate}}
\def\ENDITEMIZE{\end{itemize}}
\def\ENDENUMERATE{\end{enumerate}}

\makeatletter
\long\def\unmarkedfootnote#1{{\long\def\@makefntext##1{##1}\footnotetext{#1}}}
\makeatother

\newcommand{\sysFull}{Coalescent Computing}

\newcommand{\sysverb}{coalesce}
\newcommand{\sysAdj}{Coalescent}
\newcommand{\sysadj}{coalescent}
\newcommand{\sysNoun}{Coalescence}
\newcommand{\sysnoun}{coalescence}

\begin{document}

\date{}

\title{\Large \bf \sysFull}
\author{{\rm Kyle C.\ Hale} \\
Illinois Institute of Technology
}

\maketitle

\begin{abstract}

As computational infrastructure extends to the edge, it will increasingly offer
the same fine-grained resource provisioning mechanisms used in large-scale
cloud datacenters, and advances in low-latency, wireless networking
technology will allow service providers to blur the distinction between
local and remote resources for commodity computing. From the users'
perspectives, their devices will no longer have fixed computational power,
but rather will appear to have flexible computational capabilities that
vary subject to the shared, disaggregated edge resources available in their physical
proximity. System software will transparently leverage these ephemeral
resources to provide a better end-user experience. We discuss key systems
challenges to enabling such tightly-coupled, disaggregated, and ephemeral
infrastructure provisioning, advocate for more research in the area, and
outline possible paths forward.


\end{abstract}





\section{Introduction}
\label{sec:intro}

We envision edge deployments (i.e., 
``cloudlets''~\cite{SATYA:2019:EDGE}) that expose virtualized resources which
can \textit{transparently} augment user devices, and which can automatically scale up or
down based on available resources, user demand, user proximity (network
latency), available network bandwidth, and spot pricing. From the users'
perspective, it appears as if their device (laptop, thin client, or smart
phone) acquires increased computational power (increased memory or disk
capacity, increased CPU count, or high-end GPU) when they wander near such
a deployment, for example, into their local coffee shop\footnote{The author acknowledges
that since we are currently living in a pandemic some readers might find this particular 
example far-fetched!}. When the user leaves the area, the
resources are revoked, and the user's device appears as it did before. 

Since Satyanarayanan first laid out the basis for this vision nearly
two decades ago with Cyber Foraging~\cite{BALAN:2002:CYBERFORAGING}, hardware,
software, and networking technologies have advanced to the point where it will
soon be possible for the transparent \textit{\sysnoun{}} of disaggregated
computational resources to client machines to occur at a fine granularity and
over short time scales. We call this notion \textit{\sysFull{}}, a type of Cyber Foraging
for disaggregated hardware at the edge.

Cyber Foraging generally relies on discoverable services in
users' local environments, and on the ability to \textit{offload} application components
to remote---and generally more capable---machines~\cite{SATYA:2015:CLOUD-OFFLOAD}. This
offloading usually happens at the granularity of virtual machines~\cite{HA:2013:JIT-CYBERFORAGE};
while the end user may be unaware of the local/remote distinction in this
scenario, it is still present for the application programmer and system
software. Though applications can be automatically partitioned into
loosely-coupled components to make
them amenable to such cloud offload~\cite{CHUN:2011:CLONECLOUD, HUNT:1999:COIGN}, and while VM migration can be used to ship
applications to the cloud transparently~\cite{HA:2017:VM-HANDOFF}, we believe that there is
an opportunity to leverage the increasingly hierarchical and disaggregated
structure of cloud resources at the edge~\cite{TONG:2016:HIERARCHICAL-CLOUD, TRIVEDI:2020:EDGE-SHARING} to support
applications that are more tightly coupled, including enhanced
gaming, augmented reality (AR)~\cite{ZHANG:2017:ARNET}, virtual reality (VR)~\cite{XIE:2021:QVR}, interactive data analysis~\cite{SATYA:2019:EDGE-NATIVE}, and
IoT~\cite{ZEUCH:2020:NEBULASTREAM}.

One goal of classic distributed operating system work from the 70s and 80s
was to hide loosely-coupled distributed machines behind the illusion of
a single, logical system (a single system image~\cite{BUYYA:2001:SSI}). While commercially this did not come to pass (except for limited
components, e.g. file systems), a natural question
now arises: is it time to reconsider this aspiration for our modern computational
ecosystem?
In the datacenter, the answer seems to be \textit{yes}. High-speed
interconnects (e.g. InfiniBand) within the
datacenter have increasingly made
resource sharing between systems feasible~\cite{KALIA:2016:RDMA-DESIGN, TSAI:2017:LITE}, for example shared remote
memory~\cite{OUSTERHOUT:2010:RAMCLOUD,AGUILERA:2018:REMOTE-REGIONS,ZHANG:2015:MOJIM,
DRAGOJEVIC:2014:FARM} or remote swap~\cite{GU:2017:INFINISWAP,
AMARO:2020:FASTSWAP, RUAN:2020:AIFM-ATC}.  I/O device virtualization is
becoming more sophisticated, with efficient offload enabled by API
remoting~\cite{DUATO:2010:RCUDA, BACIS:2020:BLASTFUNCTION, SANI:2019:IODEV}, and device sharing
among tenants~\cite{YU:2020:AVA}. These trends, along with hardware proposals
for scale-out systems~\cite{LOFTI:2012:SCALEOUT-CPUS, NOVAKOVIC:2014:SONUMA,
GAVRIELATOS:2018:SOCCNUMA} and disaggregated hardware on the
horizon~\cite{ASANOVIC:2014:FIREBOX, HP-THEMACHINE,
FARABOSCHI:2015:MEMORY-CENTRIC-OS, CHUNG:2018:COMPOSABLE-SYS} point towards
a datacenter that consists of loosely-coupled, disaggregated resources. LegoOS,
a notable first step in disaggregated operating systems, embraces this view of
the datacenter while retaining Linux ABI
compatibility~\cite{SHAN:2018:LEGO-OS}, and GiantVM demonstrates how to
virtually and transparently compose datacenter
resources~\cite{ZHANG:2020:GIANTVM}.

While datacenters obviously benefit from low-latency, wired interconnects and
relatively static hardware configurations, the momentum in disaggregation is
encouraging for commodity computing at the edge as well, and we argue that there is
a ripe opportunity for OS research to make \sysFull{} a reality. 
Below we describe \sysFull{} in more detail, 
discuss some of its key research challenges, and ideas for OS design to investigate
the space.

\section{\sysFull}
\label{sec:sys}

\begin{figure}
    \centering
    \includegraphics[width=\columnwidth]{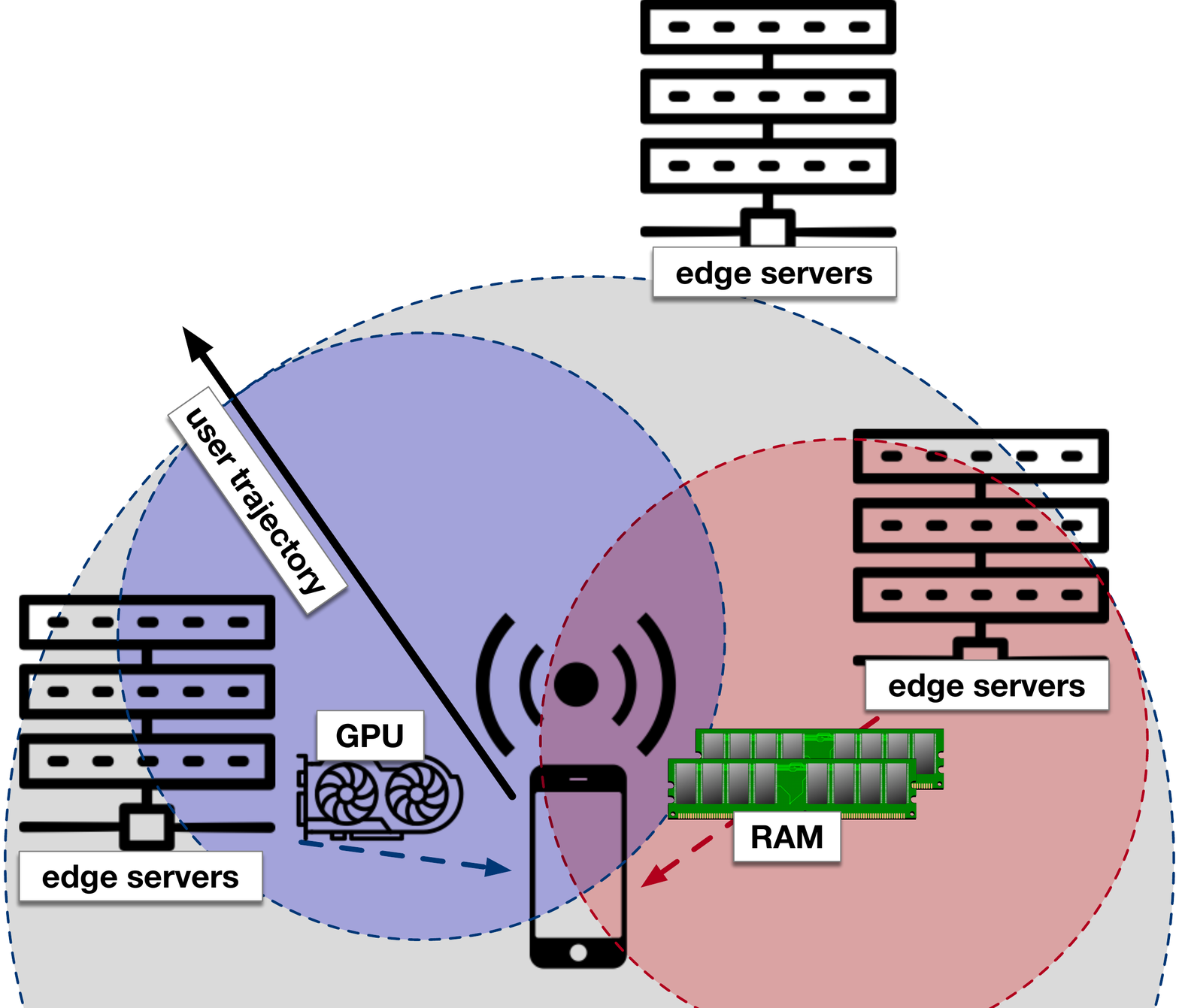}
    \caption{\sysFull}.
    \label{fig:idea}
\end{figure}

\sysFull{} can be captured succinctly with the following principle:

\begin{quote}
    \textbf{\sysNoun{} Principle:} Users' devices experience a \sysnoun{}
    of resources proportional to proximity as users move through the physical environment.
\end{quote}

Figure~\ref{fig:idea} depicts the \sysNoun{} Principle at work. The number of
resources coalesced into a user's device is inversely proportional to the user's
network distance to the hosting machines\footnote{This may or may not
correspond to physical distance, e.g. in wired settings.}.  While in most
cases those hosting machines will be stationary, in some cases, users with less
constrained devices may offer their resources to nearby users, as in
Femtocloud~\cite{HABAK:2015:FEMTOCLOUD}.  Other metrics, such as pricing,
network congestion, and power budgets for the systems hosting disaggregated
resources will also affect availability. 

\begin{figure}
    \centering
    \includegraphics[width=\columnwidth]{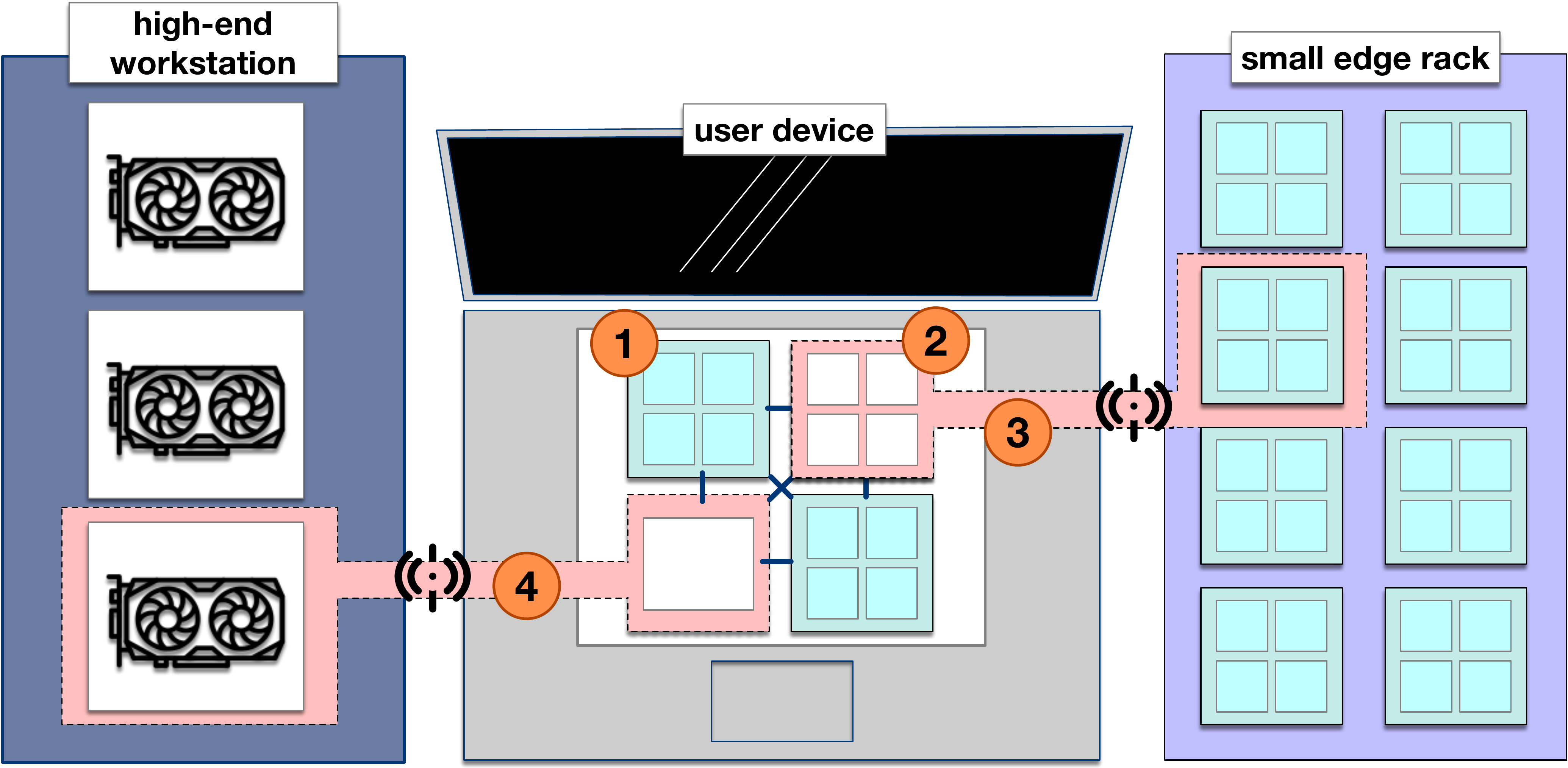}
    \caption{\sysNoun{} of disaggregated remote cores and a GPU 
    in the user's proximity.}
    \label{fig:overview}
\end{figure}

As a user navigates the physical environment, the OS on their device queries
nearby resource availability, and puts in bids for resources based on current
and historical system load.  For instance, a user that just finished a recorded
Zoom call might cause a CPU spike that triggers their OS to bid for nearby
leasable CPUs to aid in video encoding.  If no such resources are available, 
the system can either borrow resources from the traditional cloud (with a latency penalty),
or fall back to local resources. Figure~\ref{fig:overview} depicts
a scenario where a user is playing a CPU and GPU-intensive game that outstrips
the abilities of his or her laptop. The user is in nearby proximity of a high-end workstation
housing several GPUs and an edge rack that contains a collection of disaggregated
CPUs. The user sees both physical resources on the local machine (\ocirc{1}) and virtual resources from remote systems (\ocirc{2}). To accommodate system load, the device's \sysAdj{} OS transparently discovers, negotiates,
and acquires a virtual GPU \ocirc{4} and four virtual cores \ocirc{3} from the rack over
the wireless link.
This reactionary resource provisioning is reminiscent of computational
sprinting~\cite{RAGHAVAN:2012:SPRINTING, MORRIS:2018:MODEL-SPRINTING} and
JIT-provisioned Cyber Foraging~\cite{HA:2013:JIT-CYBERFORAGE}, but it happens
at the granularity of disaggregated cores and devices.
Note that a corollary of the \sysNoun{} Principle is that as users leave the
environment resources are relinquished and the system gracefully migrates any
necessary computations or state back to the local machine, or to the cloud if
WAN latencies can be tolerated.

We expect a system implementing \sysFull{} to have the following properties:

\paragraph{Transparency}

While users may set \sysnoun{} policies ahead of time, the system transparently
acquires and relinquishes resources nearby in the environment; this is a key
distinguisher from typical cloud offload. Acquisition here
does not mean sole ownership of the physical remote resource, and does not necessarily imply that the user
should be charged for it. For example, an idling user's system may make
resource reservations, but as long as the user is idle, 
the OS or monitor on the remote system is free to schedule other work.
Quality-of-Service (QoS) policies
and the
degree of sharing can be determined by providers, and will
likely change from user to user. Because users' resource acquisition policies may be at odds with one another,
and because applications have diverse requirements, policy enforcement will involve solving
a challenging, multi-objective optimization problem~\cite{FU:2021:MCIOT} on
hosts that expose resources. While there are effective techniques
for solving such scheduling problems in the datacenter---for example with recommender systems~\cite{DELIMITROU:2013:PARAGON} or
reinforcement learning and Bayesian optimization~\cite{PATEL:2020:CLITE}---in this setting
user mobility will significantly affect resource availability and users' devices may \sysverb{} resources from different providers,
rendering a centralized scheduling mechanism ineffective. One possible path forward is to combine application
profiling (informing resource \textit{requests}) with decentralized versions of ML-based schedulers (guiding
resource \textit{grants}). Thus, sets of leasable resources (servers, desktops, and possibly user devices) 
form ad hoc networks to run a distributed, \sysadj{} scheduler. 

The user can monitor currently ``attached'' resources using familiar means. For
example, \verb|/proc/cpuinfo| in a \sysAdj{} OS exposing a Linux-like interface
would include both physical
CPUs on the device and virtual CPUs coalesced from a nearby edge server.
Similarly, \verb|/proc/meminfo| would show remotely coalesced pages (though
sub-page granularity remote memory is a possibility~\cite{RUAN:2020:AIFM-ATC}).

At a surface level, the OS sees the remote CPUs (and other resources) just like
normal CPUs, and once they are properly initialized and booted, the OS can
schedule work on them.  However, the OS must take care in how it schedules work
on remote resources when applications are tightly coupled, so must have some
notion of resource localization (see Section~\ref{sec:os}). This bears some
similarity to NUMA-awareness, but is more challenging given the
inherent dynamism of resources whose \sysnoun{} depends on user proximity.

\paragraph{Performance}

Users will expect their devices to be responsive. While there is more
flexibility here than in the datacenter environment, the underlying technology
presents more challenges too (Section~\ref{sec:challenges}). In particular, as
resources attach and detach from user devices, it should not perceptibly affect
response times. 

While the single-system image abstraction is a compelling one (e.g., CPU cores
come and go as the user moves around), not all applications will want to
\textit{use} those cores, since their use comes with a latency penalty. The
system must be aware of the distinction between latency-sensitive and
throughput-sensitive workloads~\cite{SOLANTI:2020:POCLR}, and must guide
resource \sysnoun{} with that in mind. We believe that there is likely a sweet
spot for applications that thrive in a \sysadj{} setting. For applications with
components that communicate quite often (e.g., a tightly-coupled,
multi-threaded stencil code), decoupling the components will incur
a significant penalty. On the other hand, loosely-coupled applications that
perform bulk computations (e.g., rendering a single scene) can be sufficiently
handled by offloading to a distant cloud. Thus, identifying applications that
fit into this sweet spot is a primary concern. 

\paragraph{Resilience}

Though we can envision \sysFull{} extending to wired environments\footnote{For example,
inductive charging surfaces seen in some coffee shops now might one day incorporate
network interfaces.}, systems will more often
need to make do with unreliable wireless connections. A \sysAdj{} OS must
deal with dropped connections gracefully, for example using replication
and fail-over, or by periodic checkpointing.  In any case, techniques applied
to achieve resilience should avoid centralized coordination given the ephemeral
proximity of resources. However, some systems---for example edge servers housed
in a back room cabinet---will be more static by nature, will have a constant
power source, and will likely have a reliable wired connection to the internet,
and thus should be weighted more heavily when choosing coordinating nodes.

\paragraph{Customizability}

While users need not normally tend to resource \sysnoun{} policies, we
anticipate that there will arise scenarios where customization will be
advantageous.  For example, users may set a lower threshold on battery levels
at which the system discovers, negotiates, and leases resources, thus limiting
power consumption by the wireless radio and by the system itself. Even if one
user has a high-end laptop, he or she likely would not want another user
pegging one of the CPUs when the battery is on its last leg.  Other users might
prefer more detailed performance tuning, for example setting thresholds on swap
space using remote memory, capacity limits on leased resources, CPU load
thresholds for offloading, and so on.

\paragraph{Privacy and Security}

When users offload computation to cloud resources or instantiate VMs on public
infrastructure, they place some degree of trust in the provider, since they
direct the action. With \sysFull{}, a user's application may be run on untrusted
hardware, potentially divulging sensitive information. Some malicious users may
be incentivized to lease out their resources just to compromise other users' data.
Others may coalesce resources from nearby users (possibly coordinating with
other bad actors nearby) to carry out a denial-of-service attack.  Systems must
have mechanisms in place to mitigate such scenarios. This is a problem that also
plagues decentralized volunteer computing systems~\cite{KOPAL:2017:CHEAT, DURAND:2015:BITWORKER}.
Proper isolation using hardware support, virtualization, and collaborative monitoring
and reporting of bad actors can alleviate the effects of malicious behavior.

\section{Challenges}
\label{sec:challenges}

We now discuss major challenges both in hardware and in OS design
that impede progress in realizing \sysFull{}.

\subsection{Hardware} 
\label{sec:hw}

\begin{table}
    \begin{tabular}{@{}lr@{}}
    \toprule
        \textbf{Technology} & \textbf{Latency} \\
        \midrule
        SoL lower bound at 10m & 33 ns \\
        Cross-core cache-coherence & 100-200 ns~\cite{KAESTLE:2016:SMELT} \\
        soNUMA (proposed) & 300 ns~\cite{NOVAKOVIC:2014:SONUMA} \\
        Cross-socket (QPI) &  355 ns~\cite{CHOI:2016:CPU-FPGA-ANALYSIS}\\
        Inter-processor Interrupts (IPI) & $\sim$500 ns~\cite{HALE:2018:NEMO-MASCOTS} \\
        PCIe Gen 3 & 900 ns~\cite{NEUGEBAUER:2018:PCIE} \\
        InfiniBand RDMA (one-sided) & 1 $\mu$s~\cite{KALIA:2016:RDMA} \\
        WiFi 6E (reported) & $\sim$2 ms~\cite{NOKIA-WIFI6E} \\
        5G URLLC (reported) & 1 ms~\cite{FEHRENBACH:2018:5G, 3GPP-5G} \\
        5G (first-hop) &  14 ms~\cite{NARAYANAN:2020:5GPERF} \\
        Typical WiFi ($90^{th}$ \%-ile) & 20 ms~\cite{SUI:2016:WIFIPERF}\\
        \bottomrule
    \end{tabular}

    \caption{One-way latency for common and emerging networking technologies.}
    \label{tab:networks}
\end{table}

The overriding challenge for \sysFull{} from the hardware perspective will be the
performance characteristics of wireless links. Table~\ref{tab:networks} lists
single-hop latencies for various interconnects up and down the stack reported by others.
Cache line transfers on the coherence network between Nehalem cores land in the
100-200ns range, whereas high-performance InfiniBand cards are still more than
3X that latency at $\sim$1$\mu$s.  As Shan et al. have already shown in the
datacenter environment, this puts coherent resources off the table for
now~\cite{SHAN:2018:LEGO-OS}. This especially rings true for
wireless technologies (last four rows of Table~\ref{tab:networks}).  Typical
WiFi connections have reasonably low first-hop latency at 20ms, but there is
a long tail that puts the damper on deterministic performance. However,
emerging, ultra-low latency wireless standards like 5G URLCC (designed with
applications like wireless factory automation and AR in mind) and WiFi 6E bring
the latency down by an order of magnitude and are reported to reduce latency
variance significantly. Coherence will still be out of
reach, but with the right OS support we believe \sysFull{} can be realized over
these low-latency wireless links. For reference, the first row of the table shows the
speed-of-light delay at 10m, which we can view as a lower bound on the latency
of future wireless networking between edge systems. While there is much 
work on characterizing and improving the performance of wireless links, 
little has been done to guide automated decisions based on their properties. 
In particular, for a \sysadj{} system to work properly, it must be able to 
infer signal strength (and user distance) accurately in order to project the impacts
on application performance and thus guide \sysnoun{} dynamically. This is an open problem.

Unfortunately, current wireless interfaces are not suitable for operating
with disaggregated resources. Prototype systems for disaggregated hardware today
make heavy use of RDMA capabilities and fixed network latencies. 
WiFi interfaces could be optimized for \sysFull{}, for example by customizing
the wire protocol for resource acquisition, and by integrating low-power
mechanisms for resource discovery, as in BlueTooth Low Energy (BLE). These
NICs might also incorporate features we see in high-end cards today like RDMA,
atomics, and memory protection. The NICs might also be integrated near the processors to act as
a proxy socket to facilitate communication between remote resources, as in
soNUMA~\cite{NOVAKOVIC:2014:SONUMA}. 

While the OS may employ loosely-coupled monitors on remote resources
(Section~\ref{sec:os}), users may want to customize the software they run on
these resources. This will require enhanced lightweight virtualization in
wireless NICs, namely self virtualization (e.g., SR-IOV) and boot protocols that
incorporate disaggregated hardware (extended PXE). NICs on the users' systems
must coordinate with the BIOS (e.g. via a lightweight platform management
controller or a BMC) in order to keep hardware information exposed to the OS
(namely, ACPI tables that enumerate NUMA regions and processor information like
the SRAT and SLIT tables) consistent with coalesced resources.  ACPI likely
needs to be extended to support \sysFull{}, and platform hardware will need to
route the boot sequence (e.g. the SIPI and IPI sequence on x86 chips) through
something like an APICv~\cite{INTEL-APICV} rather than applying the traditional \textit{trap-and-emulate}
model.

\subsection{Software}
\label{sec:os}

A \sysAdj{} OS will need to support the following: performance,
disaggregation, resource discovery, adaptation, hardware heterogeneity, and
fault tolerance.  Several OSes from the research community support some of
these features, but not all. For example, LegoOS is the first OS designed for
disaggregated hardware~\cite{SHAN:2018:LEGO-OS}, and provides a good foundation
to build upon for \sysFull{}. The idea of stateless, loosely-coupled monitors
running on disaggregated hardware components will serve a \sysAdj{} OS as
well.  However, the LegoOS design focuses on datacenter applications, and the
ExCache-based memory management, the global resource managers, and the
InfiniBand/RDMA-based RPC will not transfer easily to a wireless
edge setting without significant hardware enhancements.

\paragraph{Performance}
To reconcile privacy and performance, users will likely want their code and
data to reside in isolated environments. This means that monitors will need to
employ very lightweight, fast-start hardware virtualization, which we have
previously shown is possible on the order of microseconds~\cite{WANNINGER:2021:VIRTINES}. 
Light-weight, virtual execution environments
will be launched on-demand to host second-level monitors from the mobile user's
system. Hardware monitors will isolate user monitors from one another.
Virtualization hardware enhancements discussed in the previous section will
make this more efficient, and a \sysAdj{} OS will likely incorporate
something like the \textit{boot drivers} used in Barrelfish/DC to account for
dynamically changing CPU information not supported in
ACPI~\cite{ZELLWEGER:2014:BARRELFISHDC}.  The CPU boot process will look much
more like the plug-and-play PCI probing process present in commodity OSes
today. For undersubscribed CPUs, the resource monitor
may use CPU hot-remove functionality to space-partition the user monitor, reminiscent of
co-kernels in Pisces~\cite{OUYANG:2015:COKERNELS}. As with Barrelfish/DC, decoupling the OS from the underlying hardware
will allow for greater flexibility with dynamic OS updates as well, as was also
demonstrated in K42~\cite{BAUMANN:2005:DYNAMIC-UPDATE}.

Performance will be mainly limited by network latency and bandwidth. A
\sysAdj{} OS will have to employ aggressive techniques to hide network
latency and variability. The OS can avoid expensive coherence traffic by
using message passing in lieu of shared memory, for example as is done in
Barrelfish~\cite{BAUMANN:2009:BARRELFISH} and LegoOS.  Serialization costs and
software overheads must also be avoided, as we are learning with disks as SSDs
become faster~\cite{LEE:2019:FAST-SSD}.  For remote memory performance, skewed access
distributions may help~\cite{GAVRIELATOS:2018:SOCCNUMA}, allowing caches to be
used to take advantage of temporal locality, but it is unlikely to produce the
same benefits we see in the datacenter. That said, similarities between users
in geographical proximity may offer hope, and the same principles that enable
CDNs will present opportunities for deduplication and sharing in edge
systems~\cite{TRIVEDI:2020:EDGE-SHARING}. 

\sysAdj{} systems will benefit from QoS policies. These
policies can be set based on provider inputs (e.g. informed by user account
balance), social credits (``how many CPU hours has the user leased out?''), current
system load, physical proximity, and the user's affinity for particular
resources (``only coalesce memory, not CPU or accelerators'').

The system should also employ best effort \sysnoun{} of resources; namely, if
network conditions are incapable of providing adequate performance, resource
negotiations should fail, and applications can run on local resources or fall back to the 
traditional cloud. This best-effort behavior has already been demonstrated for servicing I/O requests
in MittOS~\cite{HAO:2017:MITTOS}.

Generally speaking, as Schwarzkopf et al. aptly point out~\cite{SCHWARZKOPF:2013:DATACENTER-OS}, 
deterministic performance was the albatross for early distributed OSes, and we must be
mindful of the lessons learned there~\cite{VASILAKIS:2015:ANDROMEDA, TANNEMBAUM:1985:DIST-OS}.
Hardware improvements will certainly help, but exposing performance variability to the OS
is paramount for it to make acceptable decisions.

\paragraph{Heterogeneity}
Disaggregated CPUs, GPUs, FPGAs, memory, and storage will inevitably be more
heterogeneous than in a typical datacenter. A \sysAdj{} OS must handle this
heterogeneity transparently. Monitors written for different devices can expose
a unified interface, but applications must be able to run on diverse hardware,
including different ISAs, especially as competitors to x86 gain prominence.
This system might require applications be compiled into fat binaries, but
a more flexible approach would employ an intermediate representation (IR) to
dynamically adapt the application to the ISAs of nearby resources, as in
Helios~\cite{NIGHTINGALE:2009:HELIOS}.  Such a system would make judicious use
of just-in-time (JIT) compilers on edge nodes, or in cases where performance is
less critical, language VMs.  Using JIT compilation to address heterogeneity
adds another layer of complexity for performance, as it can introduce even more
variability. Managing this variability is critical for \sysFull{}, but we have
only scratched the surface of minimizing JIT compilation
latency~\cite{KRISTIEN:2019:JITLATENCY}.

\paragraph{Resource Discovery}

As users navigate the physical environment, their devices must query nearby
systems for available resources. This resource discovery process must occur
often enough to react to load spikes, but not so often as to drain device
battery and congest the local network.  The OS and hardware might employ UPnP
here~\cite{UPNP}, as in Slingshot~\cite{SU:2005:SLINGSHOT}, but the protocol
will likely need to be enhanced to include resource load information and
performance characteristics. When negotiating \sysnoun, the OS will
automatically choose a subset of nearby resources subject to user preferences
and system load.

\paragraph{Programming Model}

The \sysAdj{} OS can by default transparently migrate computation
between local and remote machines. For example, for each new vCPU
added to the system via \sysNoun{}, the OS exposes a new run queue,
e.g., over distributed shared memory. The OS can add a thread to the
remote run queue as it would on the local system, but with a performance
penalty. For example, the code in Listing~\ref{lst:worker} shows
a trivial example of creating a worker thread that processes tasks
in a shared queue. The \sysAdj{} OS is free to schedule this thread
on a remote vCPU if available. However, since the user is mobile,
that remote vCPU may disappear. The remote worker thread
might dequeue work from the shared queue then
fail, losing that work. There are of course many techniques for handling failures and
ensuring consistency in distributed systems, but in a language like
C where mutations on shared state can happen anywhere, it is quite challenging to apply these
techniques transparently.

\begin{lstlisting}[style=CStyle, caption={Creating a worker thread.}, label={lst:worker}, language=C]
static void worker() { 
    while (1) {
        work_t * work = dequeue_work();
        do_work(work);
    }
}

void main() {
    pthread_t thr;
    pthread_create(&thr, NULL, worker, NULL);
    pthread_join(thr, NULL); 
}
\end{lstlisting}

One possibility is to expose remote resources and the potential for failure
to the programmer. Listing~\ref{lst:cc-worker} shows such an example using
a CC-aware wrapper around the pthreads runtime. Here the programmer
explicitly places constraints on the remote vCPUs that the thread can run on
by specifying the maximum acceptable latency to the remote vCPU in $\mu$s. The programmer
also indicates that the system should favor remote vCPUs over local ones
when available (\verb|EAGER_REMOTE|). Finally, the programmer specifies that
when a failure is detected by the runtime, the thread should be recreated 
on the local machine after a failure is handled by user-specified code (here
the programmer provides code to recover the work queue, e.g., with a persistent
write-ahead log).

\begin{center}
\begin{lstlisting}[style=CStyle, caption={Creating a worker thread with failure recovery using an explicit CC API.}, label={lst:cc-worker}, language=C]
#include <cthread.h>
static struct cthread_attrs {
    .latency_bound = 100, // usec
    .strategy      = EAGER_REMOTE,
    .failure       = RETRY_LOCAL,
} cta; 

static int handle_failure() {
    return recover_work_queue();
}

void main() {
    cthread_t thr;
    cthread_create(&thr, &cta, fun, NULL, handle_failure);
    cthread_join(thr, NULL);
}
\end{lstlisting}
\end{center}



In many cases, it will be preferable to manage \textit{functions} running on remote resources,
rather than execution contexts. In this case, the \sysAdj{} OS can expose a function-as-a-service
(FaaS) API as well. In addition to the typical FaaS event-triggered function invocation model, 
a CC FaaS API might also allow for RPC-like, \textit{synchronous} invocations via language annotations.
For example, a programmer might specify that a function can run on remote resources by using
a \textit{virtine} (virtual subroutine)~\cite{WANNINGER:2021:VIRTINES}, as shown in Listing~\ref{lst:virtine}.

\begin{lstlisting}[style=CStyle, caption={Coalescence with a virtine.}, label={lst:virtine}, language=C]
virtine int fun() {
    do_work();
}
\end{lstlisting}

In this case, if remote resources are available, the invocation of \verb|fun| will cause a light-weight,
isolated VM (or container) to be spawned on the remote vCPU and the function will run to completion.

\paragraph{Adaptation and Fault Tolerance}

A \sysAdj{} OS will need to adapt to changing network conditions and
resource availability. Ideas from systems like Chroma apply here~\cite{BALAN:2003:TACTICS};
resources must be monitored, and application usage estimated so that
applications can scale up and down depending on what is available. The system
may leverage redundant computations across multiple resources to mitigate tail
latency and for resilience. 

To handle failures, a \sysadj{} system will likely employ replicas and
periodic checkpointing. Replication will be more challenging than in the
datacenter environment given increased mobility, but replica selection can be
informed by mobility characteristics of different systems.  A server plugged
into the wall would be a better choice for fail-over rather than a nearby
laptop. Replicas might exist in hierarchies based on the environment.  For
example, a secondary replica may be placed on the nearby edge server, and
a tertiary replica may be instantiated in the cloud with relaxed consistency.
Append-only storage can be used to persist state changes and aid in failure
recovery.

When component or connection failures occur, or when the user moves out of
range , the system must decide how to react. This will largely depend on
application resource demands and latency sensitivity. For example, when a user
training a neural network decides to move away from a resource-rich area, the
training can be shipped off to the cloud to complete. However, a user
running an immersive augmented reality application may prefer to have all
computation and state migrated back to the local device, perhaps trading off
degraded quality for responsiveness.


\section{Conclusion}
\label{sec:conc}

Several challenges remain for \sysFull{} which we do not touch on 
here, but which we do plan to investigate. 
These include the storage interface, resource naming, and
a more detailed treatment of privacy and security (e.g. authentication).

The building blocks for \sysFull{} are gradually being put in place.  We will
soon stand at the confluence of disaggregated hardware, hierarchically
distributed clouds, and ultra low-latency wireless networks. We argue that
exploring systems that support this model will not only put more computational
power at users' fingertips, but will also shed light on new avenues of systems
research.

\section*{Acknowledgements}
\label{sec:ack}

This paper would not have been possible without valuable discussions and
feedback from Conghao Liu, Brian Tauro, Nicholas Wanninger, Rich Wolski, Peter
Dinda, and Nikos Hardavellas. This work is supported by the United States
National Science Foundation via awards CNS-1718252, CNS-1763612, CNS-1730689, 
CCF-1757964, CCF-2029014, and CCF-2028958.

\balance
\bibliographystyle{plain}
\bibliography{kyle}

\end{document}